\documentclass[runningheads]{llncs}
\usepackage{amssymb}
\usepackage{algorithm}
\usepackage{algpseudocode}
\usepackage{mathtools}
\usepackage{graphicx}
\usepackage{booktabs}
\usepackage{stfloats}
\usepackage{tabularx}
\usepackage{multirow}
\usepackage{enumitem}
\usepackage{subcaption}

\def\modelName{{PASTA}}
\newcommand{\interval}[1]{\tiny{$\pm$ #1}}

\begin{document}
%
\title{\modelName: \underline{P}athology-\underline{A}ware MRI to PET Cro\underline{S}s-modal \underline{T}r\underline{A}nslation with Diffusion Models}
\titlerunning{PASTA}

%
%
\author{Yitong Li\inst{1,2},
Igor Yakushev\inst{3},
Dennis M. Hedderich\inst{4}, \\
Christian Wachinger\inst{1,2}}
%
%
\institute{Lab for Artificial Intelligence in Medical Imaging, Technical University of Munich (TUM), Germany \and
Munich Center for Machine Learning (MCML), Germany \and
Department of Nuclear Medicine, Klinikum rechts der Isar, TUM, Germany \and
Department of Neuroradiology, Klinikum rechts der Isar, TUM, Germany
\email{yi\_tong.li@tum.de}}
\maketitle              

\begin{abstract}
Positron emission tomography (PET) is a well-established functional imaging technique for diagnosing brain disorders. However, PET's high costs and radiation exposure limit its widespread use. In contrast, magnetic resonance imaging (MRI) does not have these limitations. Although it also captures neurodegenerative changes, MRI is a less sensitive diagnostic tool than PET. To close this gap, we aim to generate synthetic PET from MRI. Herewith, we introduce \modelName, a novel pathology-aware image translation framework based on conditional diffusion models. Compared to the state-of-the-art methods, \modelName~excels in preserving both structural and pathological details in the target modality, which is achieved through its highly interactive dual-arm architecture and multi-modal condition integration. A cycle exchange consistency and volumetric generation strategy elevate \modelName's capability to produce high-quality 3D PET scans. Our qualitative and quantitative results confirm that the synthesized PET scans from PASTA not only reach the best quantitative scores but also preserve the pathology correctly. For Alzheimer's classification, the performance of synthesized scans improves over MRI by 4\%, almost reaching the performance of actual PET. Code is available at~\url{https://github.com/ai-med/PASTA}.

\end{abstract}    
\section{Introduction}
To date, various tools support diagnosing brain disorders, including MRI, PET, and cognitive tests~\cite{addiagnosis}. 
Structural MRI reveals brain regional atrophy, while fluorodeoxyglucose PET tracks glucose metabolism. In Alzheimer's disease (AD) and related brain disorders, glucose uptake drops severely in certain brain regions~\cite{ad_glucose}. By sensitively reflecting functional disorders, PET has proven to have higher accuracy in early dementia detection and differential diagnosis~\cite{pet_differential,pet_acc}. 

\begin{figure*}[t!]
\vspace{-0.1cm}
    \centering
    \includegraphics[width=0.9\linewidth]{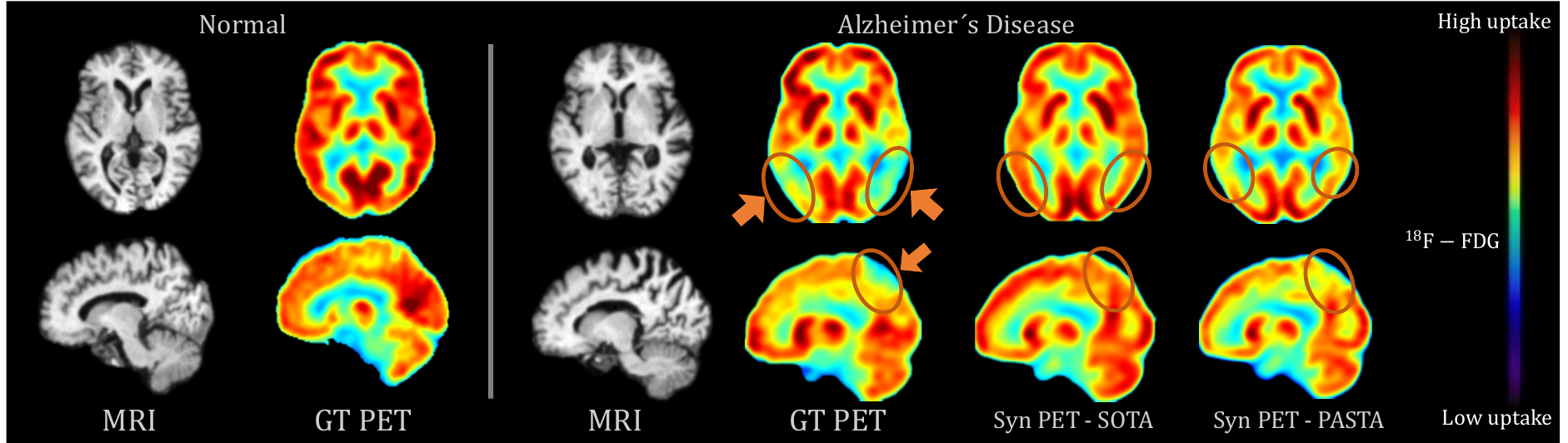}
    \caption{For Alzheimer's disease, PET shows reduced glucose uptake in the temporoparietal lobe (bottom circles), mirroring atrophy on MRI with higher sensitivity. While state-of-the-art diffusion models fail to recover such pathology in the synthesized PET, \modelName~achieves significant improvements.}
    \label{fig:intro}
    \vspace{-0.1cm}
\end{figure*}

Despite its high diagnostic value, PET is commonly not offered in most medical centers worldwide due to its high cost and high-dose radiation exposure~\cite{pet_cost}. While MRI is more accessible thanks to its non-invasive nature, it lacks the same functional insights as PET~\cite{pet_ov_mri}. 
We aim to bridge this gap by converting MRI into synthetic PET, thereby extending the reach of functional brain imaging to facilitate AD diagnosis.
Translating structural to functional imaging in medicine is challenging due to the possible inclusion of unrealistic pathology, leading to unreliable diagnoses. 
Yet, generating \emph{pathology-preserved} images remains an underexplored area with significant potential for clinical impact.  

Previous works on MRI to PET translation focus on adopting generative adversarial networks (GAN)~\cite{pix2pix,CycleGAN}. However, GANs are susceptible to mode collapse and unstable training~\cite{bau2019seeing}, which can reduce the model performance and reliability. Recent advances in diffusion models (DM)~\cite{diffusion,ddpm} introduced a new era in high-quality image generation with enhanced training stability, giving rise to advanced image translation frameworks~\cite{bbdm,Palette}. Despite these strides, existing DM-based translation methods primarily focus on preserving structural integrity, overlooking the critical aspect of pathology recovery, as shown in Fig.~\ref{fig:intro}.

We address this gap by introducing \modelName, an efficient end-to-end DM-based framework for clinically valuable pathology-aware volumetric MRI to PET translation. 
It is based on a symmetric dual-arm architecture with adaptive conditional modules for multi-modal condition integration. \modelName~also presents a memory-efficient approach for artifact-free volumetric generation and a cycle exchange consistency strategy, further lifting its generation quality.



In summary, we make the following contributions:
\begin{itemize}[label=\textbullet]
\vspace{-0.15cm}
    \item A novel end-to-end cross-modal MRI to PET translation framework based on diffusion models with volumetric generation.
    \item Adaptive normalization layers for integrating multi-modal conditions to facilitate pathology awareness.
    \item Cycle exchange consistency for effective conditional DMs training.
    \item Quantitative and qualitative experiments show that PASTA achieves low reconstruction errors and preserves AD pathology to boost diagnosis accuracy.
\end{itemize}

\subsubsection{Related Work}
\label{sec:related_works}

Previous research on cross-modal MRI to PET translation mainly focused on GAN-based methods~\cite{sketcher-refine,SCGAN,bpgan,bidirectionalGAN,ganbert,resvit,gandalf}, with innovations like the sketcher-refiner scheme~\cite{sketcher-refine}, GANDALF for MRI to PET generation in AD diagnosis~\cite{gandalf}, bidirectional GAN for 3D brain MRI-to-PET synthesis~\cite{bidirectionalGAN}, and a GAN-based residual vision Transformers for multimodal medical image synthesis~\cite{resvit}. 
Diffusion models outperform GANs in capturing complex distributions~\cite{diffusion,ddpm,DMbeatsGAN} and are emerging in medical imaging for unconditional generation and cross-contrast MRI translation~\cite{make-a-volume,generate_real_MRI_cDPM,syndiff}. Their application to MRI to PET translation, unexplored thus far, offers a promising avenue.


\begin{figure*}[t!]
\vspace{-0.3cm}
  \centering
  \includegraphics[width=\textwidth]{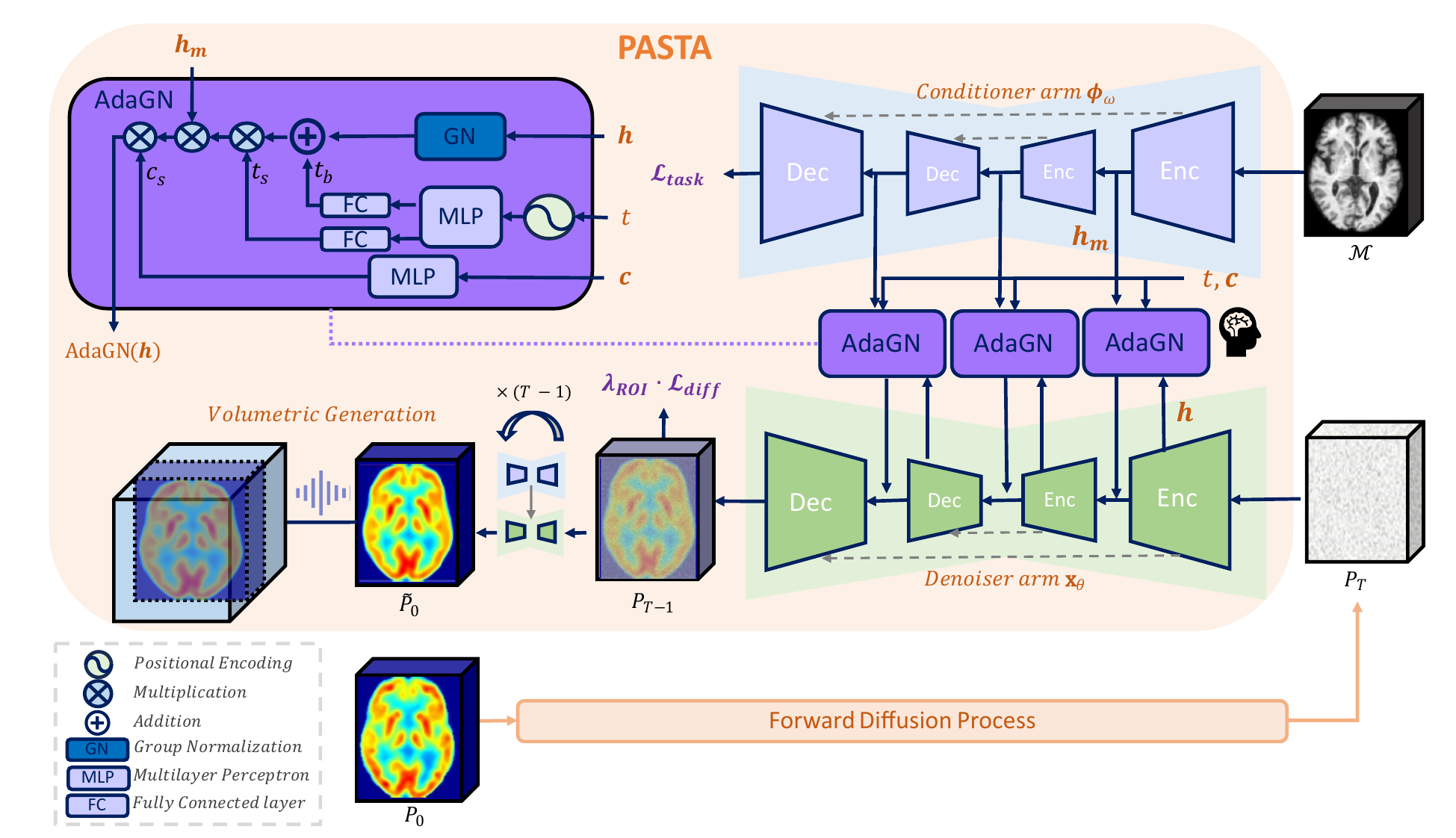}
  \caption{Overall structure of \modelName.}
  \label{fig:architecture}
\end{figure*}

\section{Proposed Method}
\label{sec:methods}

Given two datasets $\mathcal{X}_{A}$ and $\mathcal{X}_{B}$ from modalities A and B, cross-modal image translation aims to learn a mapping from A to B in a paired manner. 
In medical imaging, this process is constrained on generating images that both match the ground truth (GT) and preserve pathology evidence. 
However, current DM-based translation methods focus on style transfer and structural maintenance~\cite{ddib,diffuseIT}, which are insufficient for medical translation, as shown in Fig.~\ref{fig:intro}. 

Hence, we propose establishing a strong interaction with the input modality on the conditional denoising diffusion probabilistic models (DDPM) by integrating input features at multiple scales, giving rise to a symmetric dual-arm architecture. PASTA consists of a conditioner arm, a denoiser arm, and adaptive conditional modules, as shown in Fig.~\ref{fig:architecture}. Both arms adopt symmetric U-Net~\cite{unet}. The interaction between the two arms and the fusion of additional conditions are achieved through the adaptive group normalization layers (AdaGN)~\cite{DMbeatsGAN}. This symmetric design ensures that matching blocks across the two arms share the same spatial resolution, enabling multi-scale feature map interactions. 

\subsubsection{Conditioner Arm:}

We define our training data as $\mathcal{D}_{\text{T}} = {(\mathcal{M}^i_{\text{T}}, \mathcal{P}^i_{\text{T}})}_{i=1}^N$, which comprises $N$ pairs of MRI $\mathcal{M} \in \mathbb{R}^{H \times W \times D}$ and its corresponding PET $\mathcal{P} \in \mathbb{R}^{H \times W \times D}$. The objective is to learn a model $\boldsymbol{G}(\cdot)$ on $\mathcal{D}_{\text{T}}$, such that given any unseen MRI input $\mathcal{M}_{\text{s}} \notin \mathcal{D}_{\text{T}}$, its PET counterpart is inferred as $\mathcal{P}_{\text{s}} = \boldsymbol{G}(\mathcal{M}_{\text{s}})$. The MRI input $\mathcal{M}$ first goes through the conditioner arm to derive multi-scale task-specific representations $\boldsymbol{h}_m$, given by
 $\hat{\mathcal{M}} = \phi_\omega(\mathcal{M}; \boldsymbol{h}_m)$, 
where $\phi_\omega(\cdot)$ is the conditioner model parameterized by $\omega$. The task-specific representations $\boldsymbol{h}_m = \{\boldsymbol{h}_m^1, \dots, \boldsymbol{h}_m^n\}$ includes the intermediate feature maps from $\phi_\omega(\cdot)$ at multiple scales, with $n$ the number of residual blocks in the conditioner arm. These representations will facilitate the PET synthesis in the other arm. The conditioner arm executes predefined tasks to convert MRI into $\boldsymbol{h}_m$, including MRI reconstruction and MRI-to-PET translation, with each task dictating its own training objective. Taking MRI-to-PET translation as the predefined task, the arm is trained to minimize the pixel-level distance between the original PET and the conditioner output, using distance metric $dist(\cdot)$ like $L_1$ or $L_2$: 
\begin{equation}
    \mathcal{L}_{task}(\omega) = \mathbb{E}_{\mathcal{M}, \mathcal{P}}~dist(\mathbf{\phi}_\omega(\mathcal{M}), \mathcal{P}).
\end{equation}
We employ MRI-to-PET translation as the predefined task, as it gives the best empirical performance. The efficacy of alternative tasks is investigated in Sec.~\ref{sec:results}.

\subsubsection{Adaptive Conditional Module:}
We propose to use the adaptive group normalization layers (AdaGN)~\cite{DMbeatsGAN} to apply multi-modal conditions to the feature maps $\boldsymbol{h} = \{\boldsymbol{h}^1, \dots, \boldsymbol{h}^n\}$ in each residual block from the denoiser arm. Our AdaGN layer adapts conditions of: 1) Timestep $t$ in the diffusion process; 2) Task-specific representations $\boldsymbol{h}_m$ from the conditioner arm at corresponding scales; 3) Clinical data $\boldsymbol{c} \in \mathbb{R}^{c \times n}$ of an individual subject. Our AdaGN is defined as:
\begin{align}
    \textrm {AdaGN}&(\boldsymbol{h}, t, \boldsymbol{c}, \boldsymbol{h_m}) = \boldsymbol{c_s}(\boldsymbol{h_m} (\boldsymbol{t_s}~\textrm {GroupNorm}(\boldsymbol{h})+\boldsymbol{t_b}),
\end{align}
 where $(\boldsymbol{t}_s, \boldsymbol{t}_b) \in \mathbb{R}^{2\times c} = \textrm{MLP}(pos(t))$ is the output of a multilayer perceptron (MLP) with a sinusoidal encoding function $pos(\cdot)$, and $\boldsymbol{c}_s = \textrm{MLP}(\boldsymbol{c})$. 
These modules are used throughout the dual-arm architecture.
To enhance the accuracy and pathology preservation in generated PET scans, we incorporate available clinical data into our model, including demographic information (age, gender, education), cognitive scores (MMSE~\cite{mmse}, ADAS-Cog-13~\cite{adas-cog-13}), and AD biomarker ApoE4~\cite{apoe4}. 
Overall, the AdaGN layer fuses the multi-modal conditions including both structural and pathological evidence.



\subsubsection{Denoiser Arm}
performs the reverse process of DDPM by restoring the clean PET $\mathcal{P}_0$ from the noise. It produces $\mathcal{P}_t$ at each diffusion timestep $t$ starting from the Gaussian noise $\epsilon$ at $t = T$, conditioned on multi-modal variables via AdaGN:
\begin{equation}
  \mathcal{P}_{0:T} = \mathcal{P}_T\prod_{t=1}^T  \mathbf{x}_\theta \left(\mathcal{P}_{t-1} \mid \boldsymbol{h}_m, \boldsymbol{c}\right),
\end{equation}
where $\mathbf{x}_\theta(\cdot)$ denotes the denoising model parameterized by $\theta$. The symmetric layout of \modelName~allows the feature maps in $\mathbf{x}_\theta$ at each scale to be conditioned by an equivalent scale of the task representation from the conditioner arm, which augments the impact of the conditional modality.
To further enhance pathology awareness, we integrate MetaROIs~\cite{metaROIs} as pathology priors to guide the model to the important hypometabolic regions of abnormal metabolic changes in AD patients. MetaROIs are a set of pre-defined regions of interest based on coordinates cited frequently in other PET studies comparing AD and normal subjects. 
We convert MetaROIs into a loss weighting map $\boldsymbol{\lambda}_{R} \in \mathbb{R}^{H \times W \times D}$ and add to the denoiser arm. During training, the MetaROIs of the denoised PET images will be more penalized when they deviate from the GT PET, given by:
\begin{equation}
    \mathcal{L}^t_{diff}(\theta) = \boldsymbol{\lambda}_{R}\cdot\mathbb{E}_{\mathcal{P}_0, \epsilon} [\| \mathcal{P}_0 - {\mathbf{x}}_\theta (\alpha_t \mathcal{P}_0 + \sigma_t \epsilon, \boldsymbol{c}, \boldsymbol{h}_m) \|_2^2].
\end{equation}

\begin{figure*}[t!]
\vspace{-0.2cm}
    \centering
    \includegraphics[width=\linewidth]{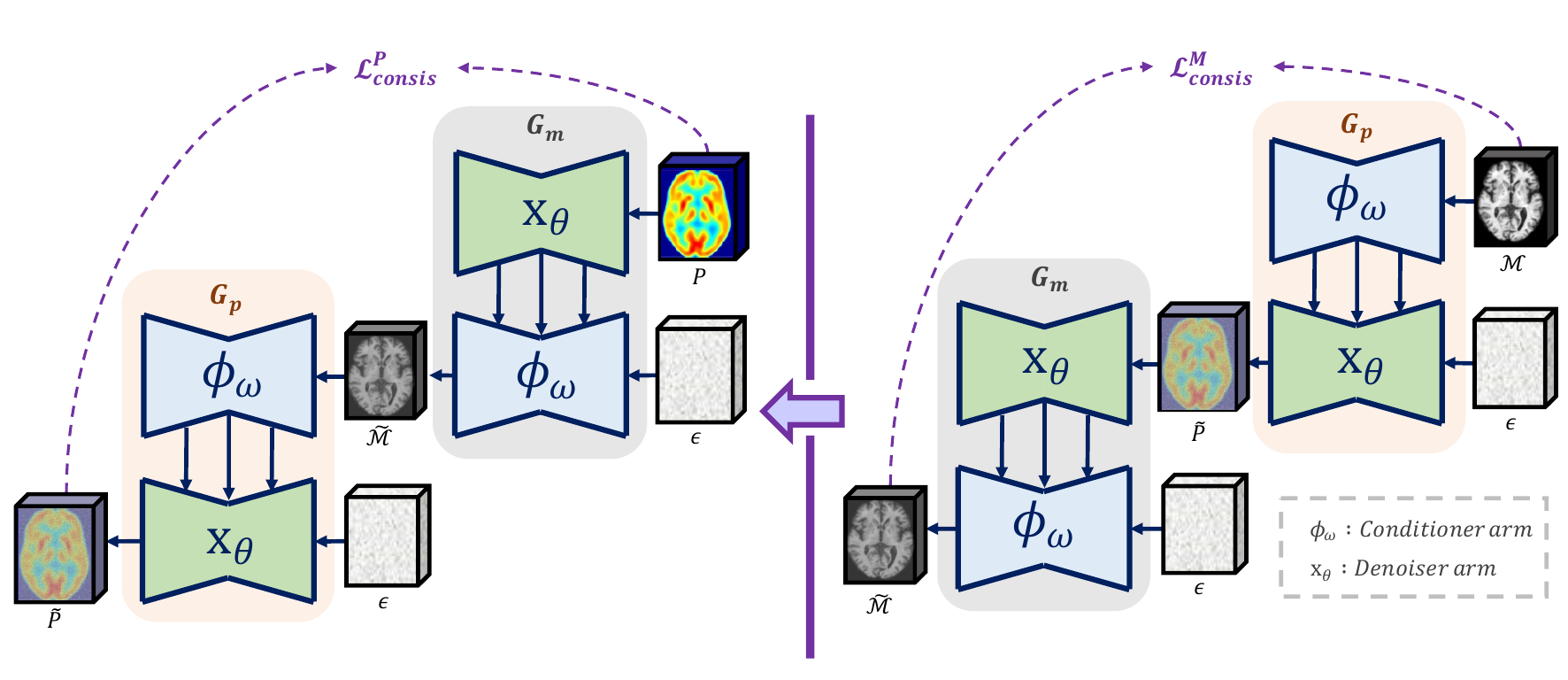}
    \caption{Cycle exchange consistency (CycleEx) strategy of \modelName.} 
    \label{fig:cycle}
    \vspace{-0.2cm}
\end{figure*}

\subsubsection{Cycle Exchange Consistency (CycleEx):}

\modelName~goes through a CycleEx strategy  shown in Fig.~\ref{fig:cycle}. CycleEx stems from the idea in~\cite{CycleGAN}, where the learned translation mapping follows cycle-consistency: for each image $\mathcal{M}^i$ from the MRI domain, given two mappings $\boldsymbol{G}_p$: $\mathcal{M} \rightarrow \mathcal{P}$ and $\boldsymbol{G}_m$: $\mathcal{P} \rightarrow \mathcal{M}$, the image translation cycle should bring $\mathcal{M}^i$ back to its original image, i.e. $\mathcal{M}^i \rightarrow \boldsymbol{G}_p(\mathcal{M}^i) \rightarrow \boldsymbol{G}_m(\boldsymbol{G}_p(\mathcal{M}^i)) \approx \mathcal{M}^i$, as forward cycle consistency. Similarly, for each image $\mathcal{P}^i$ from the PET domain, $\boldsymbol{G}_p$ and $\boldsymbol{G}_m$ should also satisfy backward cycle consistency: $\mathcal{P}^i \rightarrow \boldsymbol{G}_m(\mathcal{P}^i) \rightarrow \boldsymbol{G}_p(\boldsymbol{G}_m(\mathcal{P}^i)) \approx \mathcal{P}^i$. In the CycleEx, the mapping $\boldsymbol{G}_m$ shares the same network as $\boldsymbol{G}_p$, but only with the exchanged conditioner and denoiser arm: in the forward cycle, the conditioner arm $\mathbf{\phi}_\omega$ used to process MRI to $\boldsymbol{h}_m$ during mapping $\boldsymbol{G}_p$, will be reused for denoising to synthesize MRI in $\boldsymbol{G}_m$; the denoiser arm $\mathbf{x}_\theta$ used to synthesize PET in $\boldsymbol{G}_p$ will be reused to process PET to representations $\boldsymbol{h}_p$ in $\boldsymbol{G}_m$. The backward cycle acts similarly. Due to the symmetric nature of the two arms, this exchange can be achieved conveniently. CycleEx introduces three more conditional diffusion processes without adding additional learnable parameters. Both cycles introduce the cycle-consistency loss:
\begin{equation}
    \mathcal{L}_{cycle}(\omega, \theta) = 
    \mathbb{E}_{\mathcal{M}}~dist(\boldsymbol{G}_m (\boldsymbol{G}_p(\mathcal{M})), \mathcal{M}) + \mathbb{E}_{\mathcal{P}}~dist(\boldsymbol{G}_p(\boldsymbol{G}_m(\mathcal{P}), \mathcal{P}),
\end{equation}
where $dist(\cdot)$ can be $L_1$ or $L_2$. Such a setup enforces the information sharing between the two arms, adding additional supervision and regularization to the image translation. 
Finally, the combined training objective of \modelName~is:
\begin{equation}
    \mathcal{L} = \lambda_{task} * \mathcal{L}_{task} + \lambda_{diff} * \mathcal{L}_{diff} + \lambda_{cycle} * \mathcal{L}_{cycle},
\end{equation}
where $\lambda_{task}$, $\lambda_{diff}$, and $\lambda_{cycle}$ are constant factors.

\subsubsection{Volumetric Generation:} ~\modelName~introduces a 2.5D strategy for volumetric synthesis. A full 3D network offers inherent inter-slice consistency but is limited by high computational demands. Moreover, the shortage of paired, multimodal medical data hinders its proper training. 
Thus, we adopt 2D convolutional layers in the network with 2D slices as input, but cater to the input channel with $N$ consecutive neighboring slices of each input slice along the same axis. After training, the network produces the target slice with its $N$ neighbors. 
These neighbors are weighted by the distance to the target slice, with closer ones higher. Summing all the weighted slices and averaging overlaps for each slice position yields a consistent 3D scan.
This strategy enables the model to efficiently mitigate the slice inconsistencies introduced by the 2D network.

\section{Experiments}
\label{sec:experiments}


\noindent \textbf{Models and Hyperparameters:} We adopt U-Net for diffusion models as in~\cite{DMbeatsGAN} and DDIM sampling strategy~\cite{ddim} with timestep $T = 1000$, input neighboring slices number $N = 15$, and $\lambda_{task} = 0.1$, $\lambda_{diff} = 1$, $\lambda_{cycle} = 1$ for the training objective after an exhaustive search. Sec. A.1
reports more model parameters. 

\noindent \textbf{Datasets and Preprocessing:} We use 1,248 paired T1-weighted MRI and PET from the Alzheimer’s disease neuroimaging initiative (ADNI) database~\cite{adni}. We include the data of cognitively normal (CN, n=379), subjects with mild cognitive impairment (MCI, n=611), and Alzheimer’s disease (AD, n=257). Both modalities are co-registered with the size of $96 \times 112 \times 96$. Data are split into train/validation/test sets using only baseline visits, ensuring that diagnosis, age, and sex are balanced across sets. Sec. A.2
details on data preprocessing.

\noindent \textbf{Baselines:} Our baseline methods include Pix2Pix~\cite{pix2pix}, CycleGAN~\cite{CycleGAN}, ResVit~\cite{resvit}, BBDM, and BBDM-LDM~\cite{bbdm}. 
ResVit~\cite{resvit} is a GAN-based method integrating ResNet and ViT for medical image translation. 
Unfortunately, other GAN-based MRI to PET translation approaches do not provide open-source codes~\cite{ganbert,gandalf,sketcher-refine}. 
To ensure the inclusion of representative GAN-based methods in our comparison, we implemented the widely-used image translation techniques Pix2Pix~\cite{pix2pix} and CycleGAN~\cite{CycleGAN}.
While none of the SOTA DM-based translation models has been used for MRI to PET translation, BBDM~\cite{bbdm} stands out for its superior replicability and performance, making it our choice for adaptation and comparison.
We also include its variation BBDM-LDM based on latent diffusion models (LDM)~\cite{stable_diffusion}. The same training and evaluation data are used as in PASTA. 


\noindent \textbf{Evaluation:} A comprehensive evaluation of all methods is performed both qualitatively and quantitatively. 
We compute mean absolute error (MAE), mean squared error (MSE), peak signal-to-noise ratio (PSNR), and structure similarity index (SSIM) between the real and synthesized PET. We implement a downstream AD classification task with 5-fold cross-validation to further validate the pathology preservation. For qualitative assessment, we present our generative results with additional evaluation on 3D-SSP maps and fairness in Sec. A.4.

\begin{figure}[t!]
\vspace{-0.5cm}
    \centering
    \includegraphics[width=0.9\linewidth]{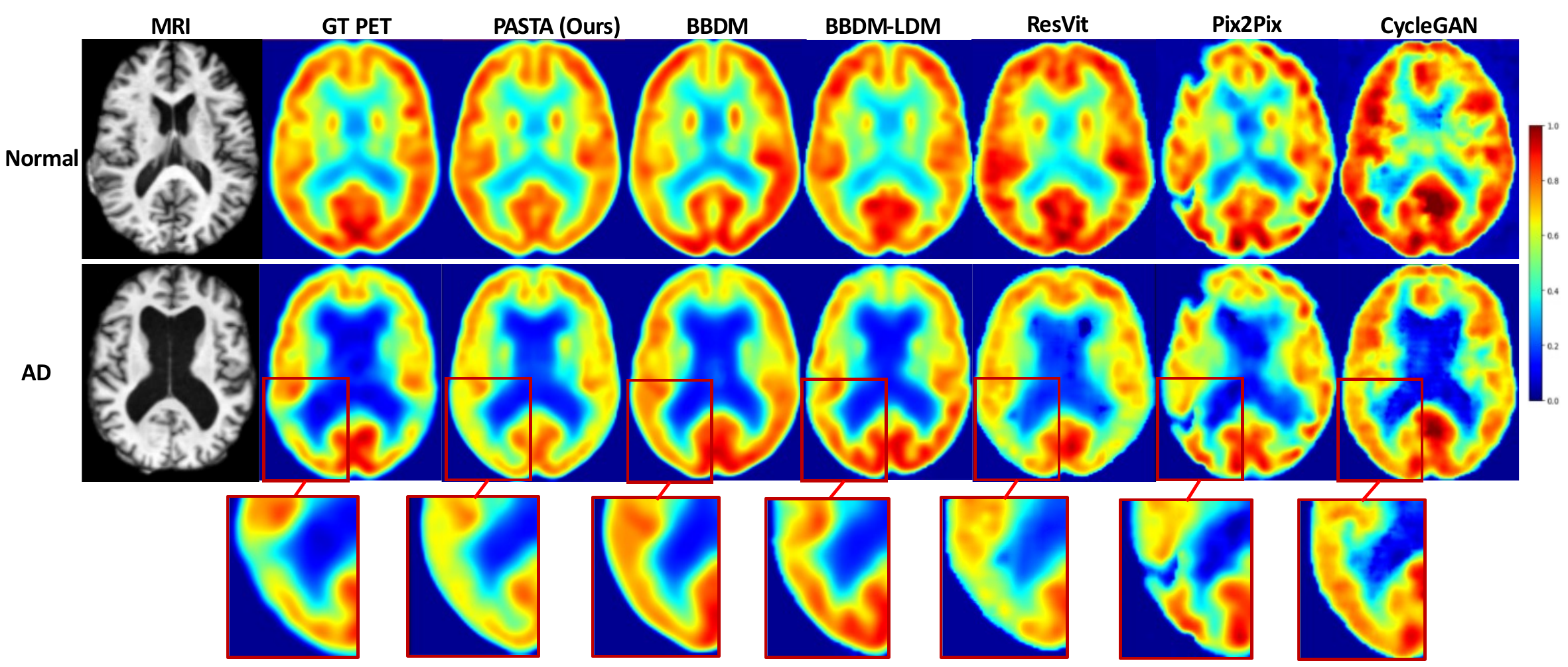}
    \caption{Qualitative results for a normal subject (top) and an AD patient (middle) with magnified specific pathology (hypometabolism in temporoparietal lobes).}
    \label{fig:bl_comp}
\end{figure}

\section{Results}
\label{sec:results}
\subsubsection{Qualitative Results and Clinical Assessment}

Fig.~\ref{fig:bl_comp} shows qualitative results for \modelName~and baseline methods.
We analyzed these results together with our clinical experts. It clearly shows that  PET generated by \modelName~have higher similarity and fidelity to the real ones than other methods. In Pix2Pix and CycleGAN, synthetic scans deviate severely from the ground truth (GT) with obvious artifacts. For AD patients, \modelName's generation accurately grasps the dropped metabolism in the temporoparietal lobe, a region highly associated with AD, as in the GT. Other DM-based models, BBDM and BBDM-LDM, recover the structure well in the generated PET but lack in transferring pathology correctly. ResVit, developed for medical translation, improves pathology awareness but generates less accurate anatomical structures. Overall, \modelName~synthesizes scans with superior consistency to the GT both pathologically and structurally.

Based on feedback from our clinical collaborators, PASTA's PET images are deemed realistic and closely comparable to actual PET. Although generated images tend to be smoother, this is not considered a drawback in clinical settings. Nuclear physicians typically apply filters on PET, and AD diagnosis does not require high-resolution edge details~\cite{pet-hypometa}.
Further, pathology is less pronounced in generated PET, which is to be expected as the synthesis relies on MRI, a modality less sensitive to functional alterations.
Yet, synthesized PETs still offer higher pathological sensitivity for AD diagnosis than corresponding MRIs.

\vspace{-0.3cm}
\subsubsection{Quantitative Results}

Tab.~\ref{tab:bl-comp} reports quantitative metrics between GT and synthesized PET for all methods.
Consistent with the qualitative results, \modelName\ generates PETs with the highest quality, owning the lowest MAE (3.45\%), MSE (0.43\%), and highest PSNR (24.59), SSIM (86.29\%). DM-based method BBDM consistently has the second-best results, with a similar performance for BBDM-LDM. However, the other GAN-based baselines, especially CycleGAN, cannot reach on-par performance. These results confirm the potential of DMs.

\begin{table}[t]
\scriptsize
 \centering
 \caption{Quantitative comparison between PASTA and the baselines.}
 {\setlength{\tabcolsep}{0.7em}
 \begin{tabular}{lcccc}\toprule
    Method & MAE(\%) $\downarrow$  & MSE(\%) $\downarrow$ & PSNR $\uparrow$ & SSIM(\%) $\uparrow$ \\\midrule
    CycleGAN \cite{CycleGAN} & 9.83 & 2.61 & 16.38 & 47.48  \\
    Pix2Pix \cite{pix2pix}    & 5.56 & 1.15 & 19.90 & 73.23 \\
    ResVit \cite{resvit} & 6.72 & 1.74 & 19.40 & 69.83 \\
    BBDM-LDM \cite{bbdm}  & 3.96 & 0.54 & 23.75 &  84.25 \\
    BBDM \cite{bbdm}  & 3.88 & 0.56 & 23.37 &  84.55 \\\midrule
    \modelName  & \textbf{3.45} & \textbf{0.43} & \textbf{24.59} &  \textbf{86.29}
    \\\bottomrule
 \end{tabular}}
 
 \label{tab:bl-comp}
 \vspace{-0.1cm}
\end{table}

\vspace{-0.3cm}
\subsubsection{Classification Results for AD Diagnosis}
To assess the synthesized 3D PET in AD diagnosis, we train AD classifiers on MRI, MRI with clinical data ($\mathbf{c}$), GT PET, and synthesized (Syn) PET, respectively, using a 3D ResNet. 
We use ResVit, BBDM, and \modelName~for the PET generation. 
To mitigate potential domain shift issues and ensure fair comparison, we train and test classifiers on images from the same source. 
Tab.~\ref{tab:classification} reports the results, and, as expected, GT PET has a higher performance than MRI. 
The results for \modelName~also surpass MRI in all metrics with an increase of over 4\%. Clinical data only boosted MRI's BACC by 2\%, suggesting that \modelName's high pathology awareness is likely due to its more effective learning of the interaction between MRI and clinical data. While the results for \modelName~in BACC are between those for MRI and GT PET, it is almost on par with GT PET in F1-Score and achieves the highest AUC. 
The results for BBDM are worse than MRI, confirming its issues with pathology transfer, consistent with Fig.~\ref{fig:bl_comp}. ResVit achieves better accuracy than BBDM, stemming from its better pathology awareness, as shown in the qualitative results, yet is still lower than for MRI.
It demonstrates the high potential of \modelName~for AD diagnosis and the necessity for pathology-aware transfer. 


 

\begin{table}[t!]
 \centering
 \scriptsize
 \caption{Results for AD classification with different input modalities.}
 {\setlength{\tabcolsep}{1.2em}
 \begin{tabular}{lccc}\toprule
    Input & BACC $\uparrow$ & F1-Score $\uparrow$ & AUC $\uparrow$  \\\midrule
    MRI   & 79.23 \interval{4.30} & 74.97 \interval{5.95} & 85.88 \interval{3.79} \\
    MRI+$\mathbf{c}$ & 81.51 \interval{3.16} & 77.83 \interval{3.83} & \underline{89.19} \interval{1.41} \\
    GT PET & {\textbf{87.02}} \interval{2.35} & \textbf{80.77} \interval{2.62} & {89.04} \interval{1.88} \\
    Syn PET (ResVit) & 78.54 \interval{4.15} & 74.41 \interval{5.28} & 81.80 \interval{4.95} \\
    Syn PET (BBDM) & 72.93 \interval{8.48} & 70.76 \interval{6.02} & 83.29 \interval{5.13} \\
    Syn PET (\modelName) & \underline{83.41} \interval{2.67} & \underline{79.98} \interval{3.51} & \textbf{91.63} \interval{2.21} \\
    \bottomrule
 \end{tabular}}
 
 \label{tab:classification}
 \vspace{-0.2cm}
\end{table}

\vspace{-0.3cm}
\subsubsection{Ablation Study}
\label{sec:ablation}

We perform ablations to verify several designs in \modelName, including CycleEx, the inclusion of pathology priors ($\boldsymbol{\lambda}_{R}$) and clinical data ($\boldsymbol{c}$), different predefined task in the conditioner arm (MRI reconstruction (M2M)) and its loss weight ($\lambda_{task}$). 
Sec. A.3
reports further ablations about condition integration and positions. 
Tab.~\ref{tab:ablation-designs} indicates that CycleEx significantly elevates generation quality, followed by the choice of predefined tasks and its loss weight. 

\begin{table}[h]
\vspace{-0.7cm}
\scriptsize
 \centering
 \caption{Ablation studies on important designs in \modelName.}
{
 {\setlength{\tabcolsep}{0.8em}
 \begin{tabular}{lcccc}\toprule
    Ablation & MAE(\%) $\downarrow$  & MSE(\%) $\downarrow$ & PSNR $\uparrow$ & SSIM(\%) $\uparrow$ \\\midrule
    w/o  CycleEx & 3.99 & 0.54 & 23.64 & 85.14 \\
    w/o $\boldsymbol{\lambda}_{R}$ & 3.67 & 0.48 & 24.19 & 85.95 \\
    w/o $\boldsymbol{c}$ & 3.57 & 0.46 & 24.39 &  86.12 \\
    \modelName~(M2M) & 3.78 & 0.50 & 23.78 & 84.77 \\
    \modelName~($\lambda_{task}$ = 1) & 3.70 & 0.49 & 24.11 & 85.71 \\
    \modelName~($\lambda_{task}$ = 10) & 3.81 & 0.48 & 24.04 & 83.24 \\   
    \midrule
    \modelName  & \textbf{3.45} & \textbf{0.43} & \textbf{24.59} &  \textbf{86.29}
    \\\bottomrule
 \end{tabular}}}
 
 \label{tab:ablation-designs}
 \vspace{-0.6cm}
\end{table}

 

\section{Conclusion}
\label{sec:conclusion}
We introduced \modelName~for translating brain MRI to PET with conditional diffusion models. Compared to current DM-based methods, \modelName\ excels in preserving both structural and pathological details in the target modality, achieved via a highly interactive dual-arm architecture. CycleEx and the volumetric generation strategy elevated its ability to produce high-quality 3D PET. 
In AD diagnosis, \modelName~reached an improved AUC over the real PET and higher accuracy than MRI. Its unique pathology awareness is likely due to the effective learning from multi-modal conditions and pathology priors, yet further research is required to explore the deeper explanation. Overall, \modelName~demonstrated high potential in bridging the gap between structural and functional brain degradation processes.


\section*{Acknowledgements}

This research was supported by the Munich Center for Machine Learning (MCML) and the German Research Foundation (DFG). 
The authors gratefully acknowledge the Leibniz Supercomputing Centre for funding this project by providing computing time on its Linux-Cluster.

%
%

\clearpage
{\bibliographystyle{splncs04}
\bibliography{supp}}

\clearpage
\appendix
\section{Supplementary Materials}
\subsection{Model Parameters and Hyperparameters}
\label{sec:models_hyperparameters}

\setcounter{table}{0}
\renewcommand{\thetable}{\thesection.\arabic{table}}

\setcounter{figure}{0}  
\renewcommand{\thefigure}{\thesection.\arabic{figure}}  


\begin{table}[h!]
\scriptsize
    \centering
    \vspace{-1cm}
    \caption{Model parameters and hyperparameters used in \modelName.}
    {\setlength{\tabcolsep}{1em}
    \begin{tabular}{llc} \toprule
        &Parameter & Value \\\midrule
        \multirow{8}{*}{Model}&Diffusion steps $T$ & 1000 \\
        &Noise scheduler $\beta$ & Cosine \\      
        &Base channels & 64 \\
        &Depth & 4 \\
        &Channel multipliers & [1, 2, 3, 4] \\
        &Attention resolution & [16, 8, 4] \\
        &Attention heads & 4 \\
        &Model size (\# parameters) & 89M \\
        
        \midrule
        \multirow{10}{*}{Training}
        &Batch size & 6 \\   
        &Optimizer & AdamW \\
        &Learning rate & $5 \times 10^{-4}$\\
        &Weight Decay & $1 \times 10^{-6}$ \\
        &Dropout & 0.0 \\
        &Training Iterations & 72K (96 hours) \\
        &EMA & 0.999 \\   
        &Diffusion loss & MAE \\
        &Predefined task loss & MAE \\
        &Hardware & one NVIDIA A100 GPU \\

        \midrule
        \multirow{6}{*}{Classification}
        &Batch size & 32 \\   
        &Optimizer & AdamW \\
        &Learning rate & $0.005$\\
        &Weight Decay & $1 \times 10^{-6}$ \\
        &Dropout & 0.2 \\
        &Training Iterations & 5K \\
        
        \bottomrule
    \end{tabular}}
    \label{tab:supple_architecture}
    \vspace{-1cm}
\end{table}

\subsection{Data Preprocessing}
\label{sec:data_process}

\begin{figure}[h!]
    \centering
    \vspace{-0.8cm}
    \includegraphics[width=0.8\linewidth]{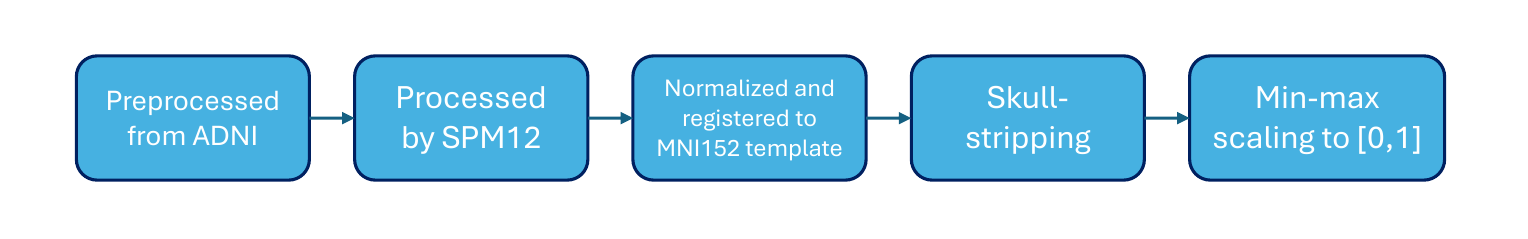}
    \caption{Data preprocessing steps for MRI and PET. Both scans have been preprocessed from the Alzheimer’s disease neuroimaging initiative (ADNI)\protect\footnotemark.}
    \label{fig:data_process}
    \vspace{-1cm}
\end{figure}
\footnotetext{http://adni.loni.usc.edu/methods}

\subsection{Additional Ablation Studies}
\label{sec:additional_ablation}
\begin{table}[h]
\vspace{-0.8cm}
\scriptsize
 \centering
 \caption{Additional ablation studies on designs in \modelName, including alternative multi-modal fusion (direct feature concatenation (ConcatFeats)), positions to integrate conditions (conditioner arm (CondCond), both arms (CondBoth)).}
{
 {\setlength{\tabcolsep}{0.2em}
 \begin{tabular}{lcccc}\toprule
    Ablation & MAE(\%) $\downarrow$  & MSE(\%) $\downarrow$ & PSNR $\uparrow$ & SSIM(\%) $\uparrow$ \\\midrule
    ConcatFeats & 3.61 & 0.48 & 24.04 & 85.00 \\
    \modelName~(CondCond) & 3.55 & 0.45 & 24.32 & 86.11 \\
    \modelName~(CondBoth) & 3.61 & 0.46 & 24.29 & 86.09 \\
    \modelName~($\lambda_{task}$ = 0.01) & 3.65 & 0.49 & 24.01 & 85.49 \\
    \midrule
    \modelName  & \textbf{3.45} & \textbf{0.43} & \textbf{24.59} &  \textbf{86.29}
    \\\bottomrule
 \end{tabular}}}
 
 \label{tab:additional-ablation-designs}
 \vspace{-0.6cm}
\end{table}

\clearpage
\subsection{Neurostat 3D-SSP Maps for PET Scans}
\label{sec:neurossp}

\begin{figure}[h]
\vspace{-0.1cm}
\centering
\begin{subfigure}[b]{0.47\textwidth}
    \includegraphics[width=\textwidth]{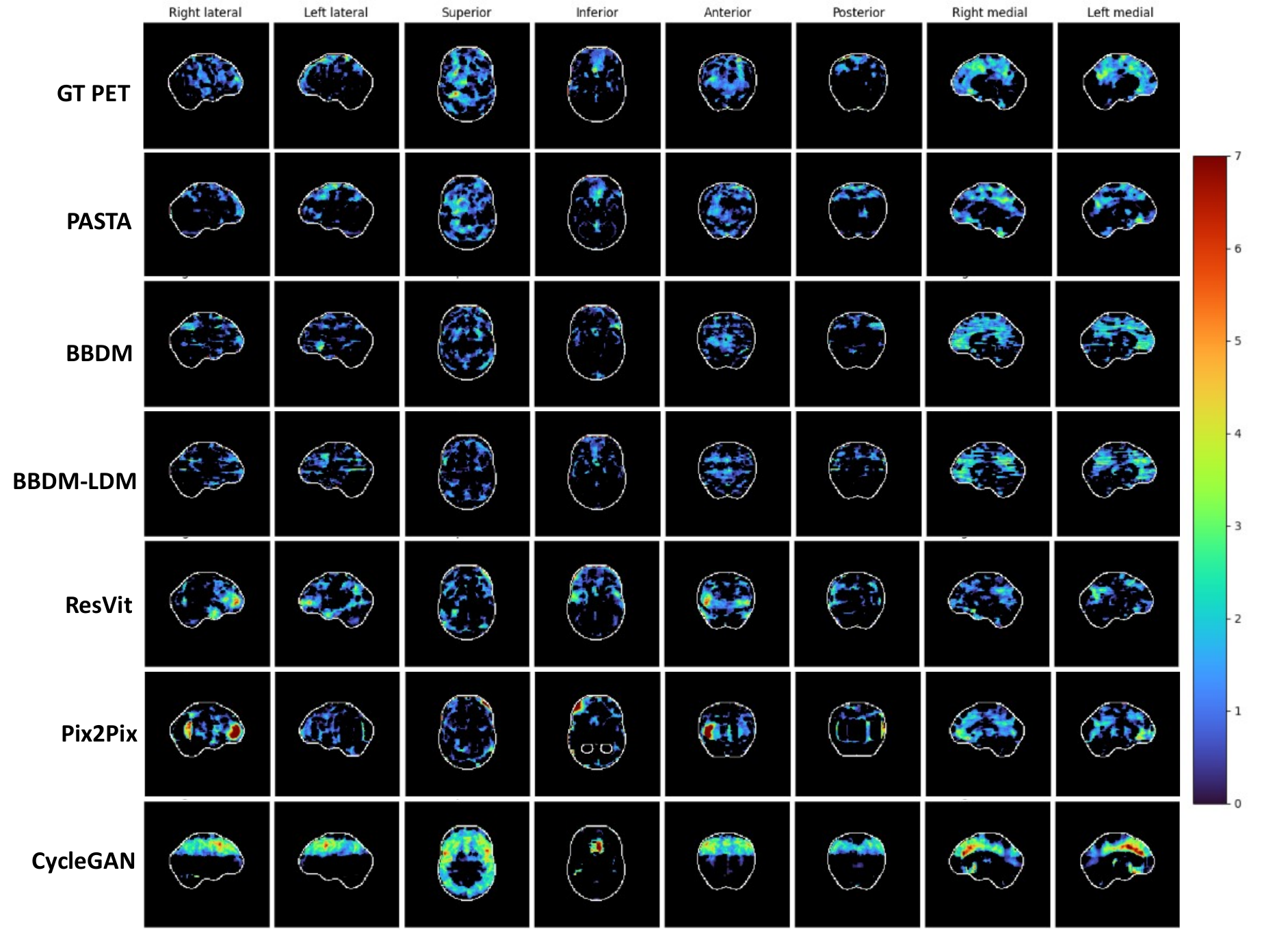}
    \caption{Healthy subject}
    \label{fig:neurostat_ssp_cn}
\end{subfigure}
\begin{subfigure}[b]{0.47\textwidth}
    \includegraphics[width=\textwidth]{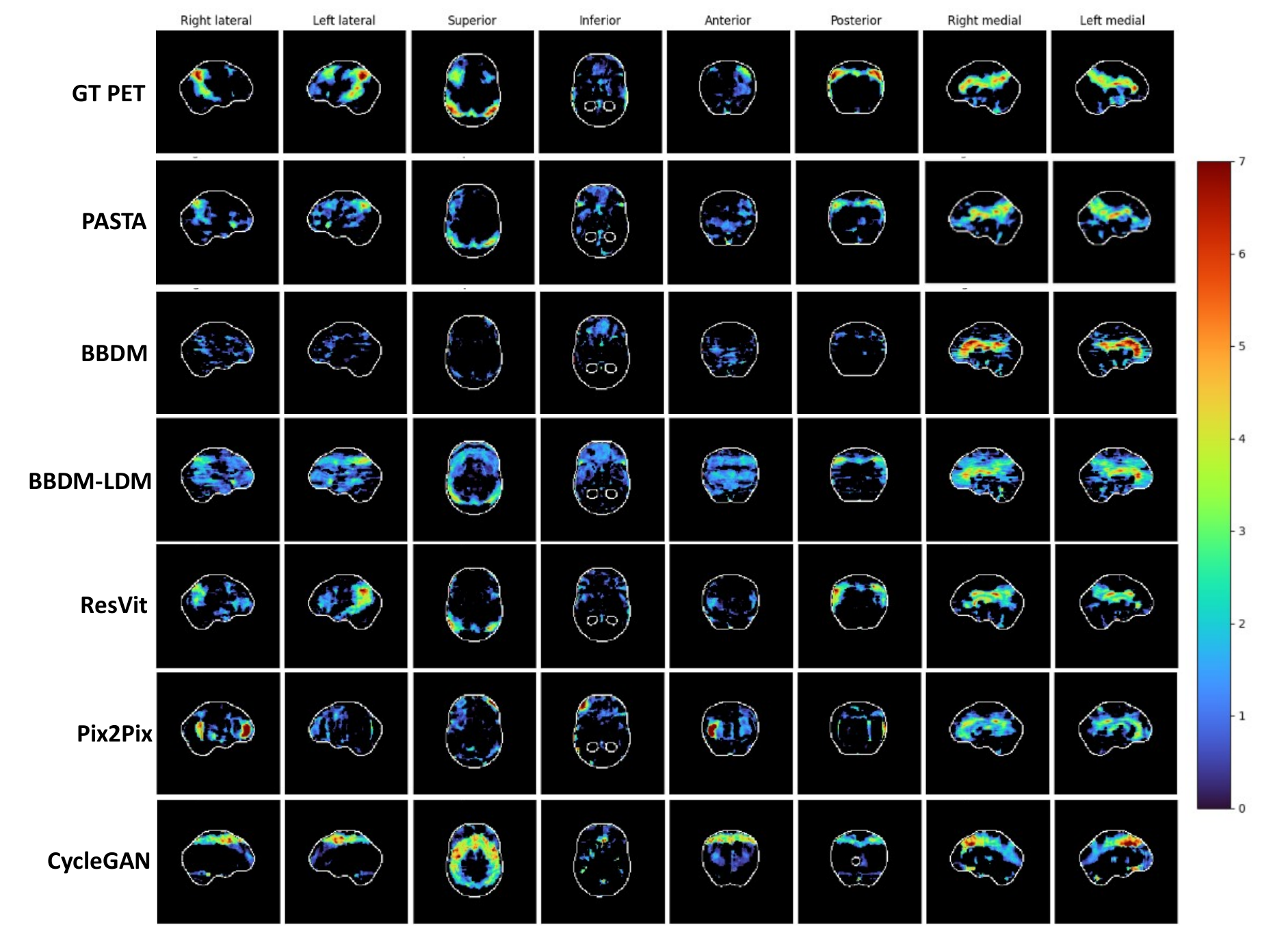}
    \caption{Alzheimer's patient}
    \label{fig:neurostat_ssp_ad}
\end{subfigure}
\vspace{-0.2cm}
\caption{
Neurostat\protect\footnotemark~3D-SSP, a statistical brain mapping technology, helps brain disorders diagnosis through PET by comparing cortical metabolic activity between patients and healthy controls, visualized as Z-score maps on the brain surface. We employ 3D-SSP to assess the accuracy of synthesized PET, including those from \modelName~and baseline methods. Results reveal \modelName~aligns closely with GT metabolic patterns in both healthy (left) and AD (right) subjects, showcasing its superior pathology preservation. Other models fail to recover the pathology evidence correctly and sometimes introduce non-existent abnormality.
}
\label{fig:neurostat_ssp}
\vspace{-0.8cm}
\end{figure}
\footnotetext{https://neurostat.med.utah.edu/}

\subsection{Fairness Evaluation}

\begin{table}[h!]
\scriptsize
    \centering
    \vspace{-0.8cm}
    \caption{We demonstrate the MAE of the synthesized PET from test samples compared to the real PET across demographics. The results indicate that the errors across different groups exhibit minimal variance and the differences are statistically insignificant (after comparing each group against the remaining samples, using the Wilcoxon rank-sum test, and applying a Bonferroni correction for multiple testing, all the p-values exceed the threshold of 0.05). Thus, it suggests that \modelName~demonstrates a uniform performance, maintaining equitable accuracy across all examined demographic categories.}
    {\setlength{\tabcolsep}{1.5em}
    \begin{tabular}{ccc} \toprule
         Demographics & Groups & MAE (\%) $\downarrow$ \\\midrule
         \multirow{4}{*}{Age} & $<$ 60 & 3.55 \interval{0.55} \\
          & 60 - 70 & 3.30 \interval{0.36} \\
          & 70 - 80 & 3.44 \interval{0.50} \\
          & $>$ 80 & 3.66 \interval{0.67} \\\midrule
        \multirow{2}{*}{Gender} & Male & 3.52 \interval{0.57} \\
         & Female & 3.35 \interval{0.41} \\\midrule
        \multirow{3}{*}{Diagnosis} & CN & 3.31 \interval{0.44} \\
         & AD & 3.59 \interval{0.46} \\
         & MCI & 3.47 \interval{0.56} \\\midrule
        \multicolumn{2}{c}{Total} & 3.45 \interval{0.51} \\
       \bottomrule
    \end{tabular}}
    \label{tab:fairness}
    \vspace{-0.5cm}
\end{table}

\end{document}